\documentclass[twoside,10pt,a4paper]{newFNLstyle}
\usepackage{graphics}
\usepackage{cite}

\def\w0{\omega_0}

\begin{document}

\volnumpagesyear{5}{2}{000-000}{2005}
\dates{}{}{}

\title{A SIMPLE NOISE MODEL WITH MEMORY FOR BIOLOGICAL SYSTEMS}

\authorsone{O. Chichigina}
\affiliationone{Moscow State University, Physics Department\\
Leninskie Gory, 119992, Moscow, Russia}
\mailingone{chichigina@hotmail.com}

\authorstwo{D. Valenti$^{\circ}$ and B. Spagnolo$^{\dagger}$}
\affiliationtwo{INFM and Dipartimento di Fisica e Tecnologie
Relative, Group of Interdisciplinary Physics\footnote {Electronic
address: http://gip.dft.unipa.it},} \mailingtwo{Universit\`a di
Palermo, Viale delle Scienze pad. 18,
I-90128 Palermo, Italy \\
$^{\circ}$valentid@gip.dft.unipa.it $^{\dagger}$spagnolo@unipa.it}

\title{A SIMPLE NOISE MODEL WITH MEMORY FOR BIOLOGICAL SYSTEMS}

\maketitle

\markboth{O. Chichigina, D. Valenti and B. Spagnolo}{Noise model
for biological systems}

\pagestyle{myheadings}



\begin{abstract}
A noise source model, consisting of a pulse sequence at random
times with memory, is presented. By varying the memory we can
obtain variable randomness of the stochastic process. The delay
time between pulses, i. e. the noise memory, produces different
kinds of correlated noise ranging from white noise, without delay,
to quasi-periodical process, with delay close to the average
period of the pulses. The spectral density is calculated. This
type of noise could be useful to describe physical and biological
systems where some delay is present. In particular it could be
useful in population dynamics. A simple dynamical model for
epidemiological infection with this noise source is presented. We
find that the time behavior of the illness depends on the noise
parameters. Specifically the amplitude and the memory of the noise
affect the number of infected people.
\end{abstract}

\section{Introduction}

Many studies have reported occurrence of fluctuations and noise in
biological systems and the role played by the noise, always
present in natural systems \cite{Sci99,Bjo01,Eco01}. The
stochastic nature of living systems and the related noise-induced
effects ranges from bio-informatics \cite{Bla03,Ozb02}, population
dynamics \cite{Tur00,Spa03,Val04,Spa04}, to virus dynamics and
epidemics \cite{Tuc98,Gie00}. Noise existing in biological systems
is due to environmental fluctuations and it is usually taken into
account as a multiplicative white noise \cite{Spa02}. The white
noise is usually defined as a continuous stochastic process with
zero mean and correlation function $\langle \xi \xi _{t}\rangle
=C\delta (t), $ where $C$ is the noise intensity. However when we
reproduce this noise in discrete time with step $\tau $ for
computer simulations, we consider rectangular pulses with width
$\tau$ and some distribution of probabilities for their heights.
In this case the randomness is in the amplitude of noise pulses.
We call this noise a discrete white noise source. In this paper we
will consider both continuous and discrete noise sources. Starting
from Markovian white noise we will introduce some memory obtaining
a quasi-periodical process, which is strongly NonMarkovian.

In some population dynamics models, the noise must be positive at
all times and often the randomness is due to stochastic natural
events, that can be represented as pulses occurring at random
times. We consider therefore a white noise source as a series of
positive pulses at random times. To obtain zero average, if it is
necessary, we can subtract the positive mean value.

In  our model the discrete white noise at $i$-th step can be
expressed as

\begin{equation}
\xi_i= \frac{f}{\tau} \thinspace \vartheta (\nu \tau-\eta_i),
\label{noise1}
\end{equation}
where $f/\tau$ is the amplitude of pulses, which we set constant
for simplicity, $\eta_i$ is a random value, distributed
homogeneously in the interval $[0,1]$, $\vartheta (x)$ is the
theta function ($\vartheta (x\ge 0) =1, \; \vartheta (x<0)=0$),
$\nu$ is the probability per unit time to have a pulse and $\nu
\tau $ is the probability per one step to have a pulse. To obtain
a time series of noise we extract random values $\eta_i$ at every
time step $\tau$. The average distance between pulses $\zeta$, can
be considered as an effective period $\langle \zeta \rangle
=T=1/\nu$ (with $T\gg \tau$). The noise average is $\langle \xi
\rangle =f/T$, and the correlation function is different from zero
only during the step $\tau$. From formal point of view this noise
is therefore white.

The continuous white noise is obtained in the limit $\tau \to 0$.
This noise is usually expressed as a sequence of $\delta $-shape
pulses at random times $t_j$ \cite{Str63}

\begin{equation}
\xi_j=\thinspace f \thinspace \sum\limits_{j} \delta (t-t_j).
\label{pulse_noise}
\end{equation}
where

\begin{equation}
t_j=t_0+\zeta _1+...+\zeta _j, \label{times}
\end{equation}
with $t_0$ the initial time, and $\zeta_j$ the random time
distances between neighboring pulses. The probability distribution
is $w(\zeta )= \nu \exp(-\nu\zeta)$, where $\zeta$ is distributed
in the interval $[0,\infty]$. Of course the effective period
$T=\langle\zeta\rangle $ and the mean value $\langle \xi \rangle$
are the same as for discrete noise, and here we consider them as
constant parameters. We will use both continuous
(\ref{pulse_noise}) and discrete (\ref{noise1}) descriptions of
the noise.

\section{Noise with memory}

For some biological applications should be interesting to consider
a noise source with some delay $\zeta _0$ after each pulse. The
system has memory during time $\zeta _0$ and the next pulse is
forbidden during this time. This process is suitable to obtain
noise sources with varying degree of randomness, ranging from
white noise, as we defined it above ($\zeta _0=0 $), to
quasi-periodical process ($\zeta _0 \simeq T$). So the probability
distribution of random time distances $\zeta_j$ is

\begin{equation}
w(\zeta)=pe^{p(\zeta _0-\zeta)},\label{prob_distr}
\end{equation}
where $\zeta$ is distributed in the interval $[\zeta _0,\infty]$,
$p$ is the probability per unit time to have a pulse. If $\zeta
_0=0$, then $p=\nu$. The first moment of $\zeta$ is therefore
easily derived as

\begin{equation}
\langle\zeta\rangle = \zeta_0+\frac{1}{p}.
 \label{period}
\end{equation}
For fixed effective period $\langle\zeta\rangle$, the probability
density $p$ increases as memory $\zeta_0$ increases. This
probability is equal to $1/\tau$, when we have periodical process
in discrete time ($\zeta _0 \to \langle \zeta \rangle -\tau$). For
continuous time, the probability of pulse per unit time $p \to
\infty$, when $\zeta _0\to T.$

For the discrete process the standard deviation of the noise is
the weighted sum of this quantity during delay (which is zero) and
that after delay $\tilde \sigma^2$, with the related
probabilities. All parameters after delay we will mark by tilde.
So we obtain

\begin{equation}
\sigma^2=\frac{\zeta_0}{T} \times 0+ \frac{T-\zeta_0}{T} \times
\tilde \sigma^2. \label{var1}
\end{equation}
For the variance after delay we have

\begin{equation}
\tilde \sigma^2 =\langle \tilde \xi^2 \rangle -\langle \tilde \xi
\rangle ^2= \left( \frac{f}{\tau} \right) ^2 p\tau -\left(
\frac{f}{\tau} p\tau \right) ^2, \label{var2}
\end{equation}
where $\langle \tilde \xi \rangle$ and $\langle \tilde \xi^2
\rangle$ are taken for the time after delay. Substituting
(\ref{var2}) into (\ref{var1}) and taking into account
(\ref{period}) we obtain

\begin{equation}
\sigma^2=\frac{T-\zeta_0}{T} p\tau \left( \frac{f}{\tau} \right)
^{2}(1-p\tau) = \frac{(T-\zeta_0-\tau )}{T\tau (T-\zeta_0)} f^{2}.
\label{var3}
\end{equation}
So we have an increasing $\sigma^2$ as the memory of the system
decreases. The standard deviation achieves its maximum value
$\sigma^2 = ((T-\tau)/(T^2 \tau))f^2 \simeq f^{2}/\tau T$ for
$\zeta_0 = 0$. When the memory is maximum $\zeta_0 = T-\tau $, the
process is not random, so $\sigma^2 = 0$.

For the continuous noise process it is useful to calculate the
spectral density. The characteristic function for $\zeta$ is the
Fourier transform of the probability density $w(\zeta)$

\begin{equation}
\Theta (\omega)=pe^{p\zeta _0} \int\limits_ {\zeta _0}^{\infty}
e^{-p\zeta +i\omega \zeta } d\zeta  =  \frac{pe^{i\omega \zeta
_0}}{p-i\omega}.\label{char1}
\end{equation}
By using the relation between the spectral density of the process
$\xi$ and the characteristic function of $\zeta$, derived in ref.
\cite{Str63}

\begin{equation}
S[\xi-\langle\xi\rangle , \omega]=\frac{2 \left (1-|\Theta
(\omega)|^2\right)}{ |1-\Theta (\omega)|^2}, \label{char2}
\end{equation}
we get

\begin{equation}
S[\xi-\langle\xi\rangle , \omega]=\frac{2 \omega^2(1-\zeta
_0)^2}{2-2\cos(\omega \zeta_0)+2(1-\zeta _0)\omega \sin(\omega
\zeta_0)+\omega^2 (1-\zeta _0)^2}, \label{char3}
\end{equation}
where we use $T = <\zeta> = 1$.
\begin{figure}[htbp]
\centering{\resizebox{9cm}{!}{\includegraphics{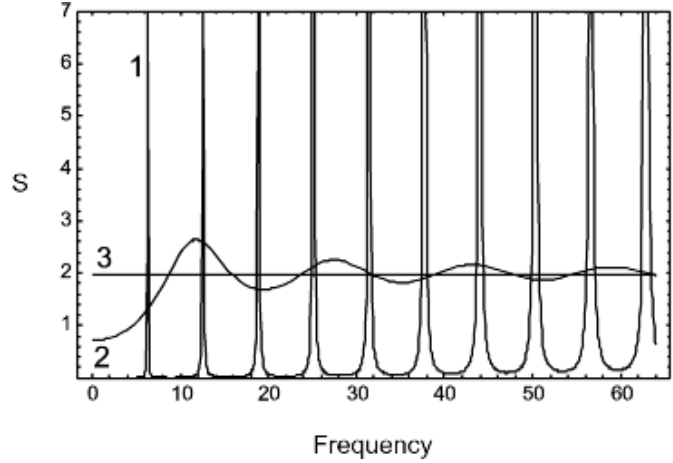}}}
\caption{Power spectrum of the process $\xi(t)$ for three values
of the memory $\zeta_0$: curves 1, 2 and 3, correspond
respectively to $\zeta_0 = 0.99, 0.3, 0.01$.} \label{power
spectrum}
\end{figure}
The power spectrum of the noise $\xi(t)$ is shown in
Fig.~\ref{power spectrum}, for three values of memory parameter,
namely $\zeta _0= 0.99, 0.3, 0.01$. We see that at $\zeta _0
=0.99$ the spectrum is a series of $\delta$-function-like pulses
with harmonics at $\omega_n \simeq 2 \pi n$. For $\zeta _0=0.01$
the spectrum density is almost constant, and corresponds to white
noise. For all cases, the spectrum density is quasi-periodical
with distance between maxima equal to: $\omega_n -
\omega_{n-1}\simeq 2\pi /\zeta _0.$

\section{Some possible applications}

The pulse noise model can be useful to describe some different
physical and biological systems, where random processes with delay
are present. It can be useful to describe and to understand the
behavior of an epidemiological infection inside a population with
a big number of individuals. In particular the number of contacts
between ill and healthy people can be modelled as our noise with
memory, where the delay describes the incubation period of the
illness. So we can compare development of illnesses with different
incubation periods.

Another possible application is the neuronal response to an
electrochemical pulse. In fact the response time of a neuron,
activated by a pulse, can depend on temperature and other
parameters. Moreover our pulse noise model allows to regulate the
periodicity, and this should play an important role in the neural
network field.

The pulse noise source can also represent an alternative way to
describe population dynamics of interacting species, previously
modelled using stochastic resonance phenomenon \cite{Val04}. This
should be very useful when a periodicity, connected with
environmental parameters, is present.

\section{Simple dynamics in the presence of noise with memory}

\noindent
 We start to analyze the role of the noise with memory
considering a simple system described by the equation

\begin{equation}
\frac{dI}{dt}=\xi(t)\thinspace I + a\thinspace I,
\label{infection}
\end{equation}
where $\xi(t)$ is the the noise with memory, $a$ is a negative
constant, and $I(t)$ can be a species density in population
dynamics or the number of infected people in epidemics. We
interpret this stochastic differential equation in the Ito sense.
This equation has been previously investigated in refs.
\cite{Zel87,Sok03}.

\begin{figure}[htbp]
\centering{\resizebox{13cm}{!}{\includegraphics{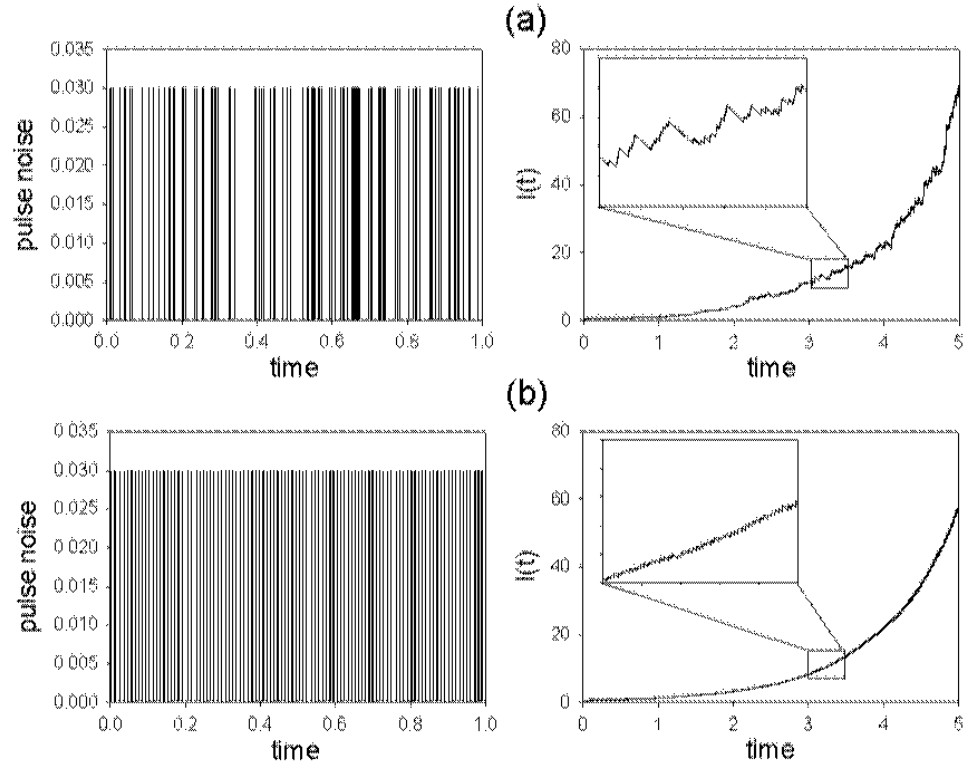}}}
\vskip-0.3cm
\caption{Pulse noise $\xi(t)$ and corresponding time
series of the number of infected people $I(t)$ vs time, for two
values of the delay $\zeta_0$. The values of the parameters are:
$T=\langle\zeta \rangle=10 \tau$, $\tau=10^{-3}$, $f=3\cdot
10^{-2}$, $I(0)=0.5$, and $a=-2$. (a) white noise source, with
$\zeta_0=0$; (b) correlated noise source, with $\zeta_0=8 \tau$.}
\vskip-0.3cm \label{time_series_1}
\end{figure}
\begin{figure}[htbp]
\centering{\resizebox{13cm}{!}{\includegraphics{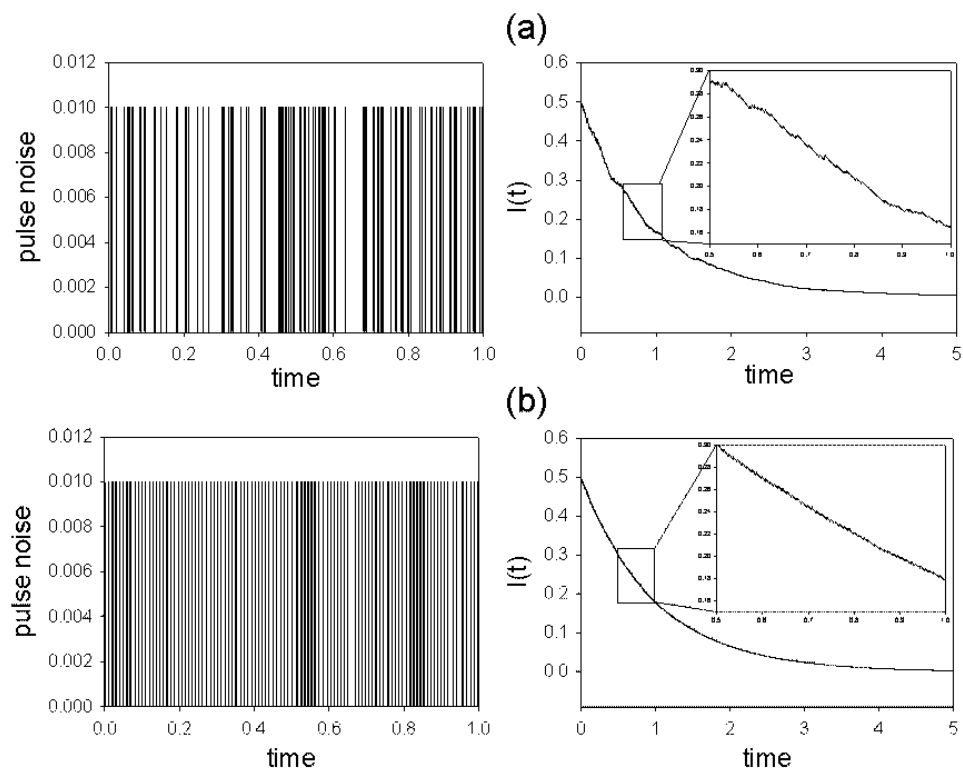}}}
\caption{Pulse noise $\xi(t)$ and corresponding time series of the
number of infected people $I(t)$ vs time, for two values of the
delay $\zeta_0$. Here $f=10^{-2}$, and all other parameter values
are the same of Fig.\ref{time_series_1}. (a) white noise source,
with $\zeta_0=0$; (b) correlated noise source, with $\zeta_0=8
\tau$.}
\label{time_series_2}
\end{figure}
\begin{figure}[htbp]
\centering{\resizebox{9cm}{!}{\includegraphics{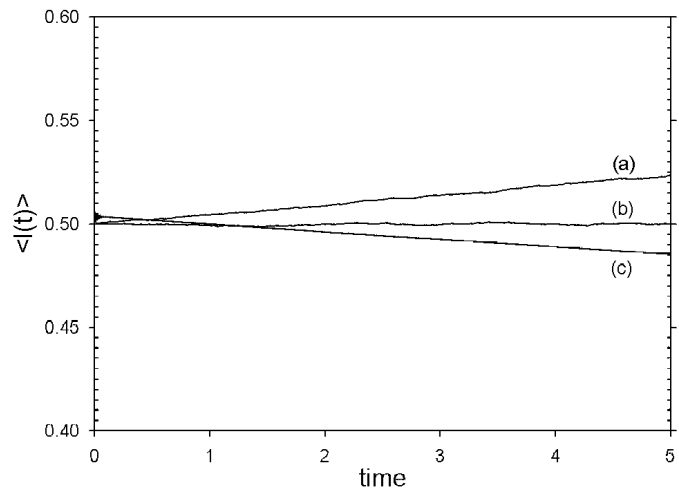}}}
\caption{Time behavior of the ensemble average of the number of
infected people $I(t)$ series, for different values both of the
delay $\zeta_0$ and the noise amplitude $f$. All the other
parameter values are the same of Fig.\ref{time_series_1}. (a)
white noise source ($\zeta_0=0$) with $f=2.01\cdot 10^{-2}$; (b)
white noise source ($\zeta_0=0$) with $f=2.00\cdot 10^{-2}$; (c)
correlated noise source with $\zeta_0=8 \tau$ and $f=2.01\cdot
10^{-2}$. The number of numerical realizations is $10^4$.}
\label{time_series_3}
\end{figure}

\noindent
The formal solution of Eq.~(\ref{infection}) is

\begin{equation}
I(t) = I(0) exp \left[at + \int_0^t \xi (t')dt' \right].
 \label{sol}
\end{equation}
By numerical integration of Eq.~(\ref{infection}) we find that
 the noise amplitude parameter $f$ has a
critical value $f_c$. For $f > f_c$ an instability occurs and the
illness exhibits a divergent behavior. By setting constant the
value of the average period $T = \langle\zeta\rangle =10 \tau$,
time series of the infection density are obtained for different
values of the delay $\zeta_0$, which represents the time before a
new pulse can occur. We report the results in
Fig.~\ref{time_series_1}, together with the pulse noise source.
The parameter setting is: $I(0)=0.5$, $f=3\cdot 10^{-2}$, $a=-2$
and $\tau =10^{-3}$. In Fig.~\ref{time_series_2} we show the same
quantities $\xi(t)$ and $I(t)$, for the same parameter setting
with $f=10^{-2}$.
In both figures an exponential behavior of $I(t)$ appears with
fluctuations more pronounced, when the noise source is white
($\zeta_0=0$) (see Figs.~\ref{time_series_1}a and
\ref{time_series_2}a). For correlated noise we have a
quasi-periodical process. The exponential behavior is decreasing
(see Fig.~\ref{time_series_1}b) or increasing (see
Fig.~\ref{time_series_2}b) depending on the value of parameter $f$
with respect to the critical value $f_c$. We observe clearly that
the noise memory affects directly the fluctuations of the number
of infected people $I(t)$. Using noise with zero memory we find
that the critical value is $f_c = 2 \cdot 10^{-2}$. The noise
memory also affects the exact value of $f_c$. In particular, when
$f$ is near the critical value with zero memory, the time behavior
of the average number of infected people changes from increasing
to decreasing exponential behavior, depending on the value of the
noise memory. In Fig.~\ref{time_series_3} we report the time
behavior of the average ensemble $<I(t)>$ for different values of
noise parameters $f$ and $\zeta_0$. Namely for (a) $\zeta_0=0$,
$f=2.01 \cdot 10^{-2}$, (b) $\zeta_0=0$, $f=f_c=2.00\cdot
10^{-2}$, and (c) $\zeta_0=8\tau$, $f=2.01\cdot 10^{-2}$. The
average is performed on $10^4$ numerical experiments. We see that
for $f = 2.01\cdot 10^{-2} \simeq f_c$ (critical value for white
noise with zero memory), by increasing the noise memory from zero
to $\zeta_0 = 8 \tau$, an instability-stability transition occurs.
The average number of infected people changes from increasing ((a)
in Fig.~\ref{time_series_3}) to decreasing ((b) in
Fig.~\ref{time_series_3}) exponential behavior.

\section{Conclusions}
The noise with memory described above is a very useful tool for
modelling quasi-periodical processes in nature. The parameter of
periodicity allows to distinguish deterministic and random
processes, and to adjust model to a real process accordingly. When
this noise is included into the equation as a multiplicative
noise, we can see qualitatively different behavior of the system
in dependence on the noise amplitude and the periodicity parameter
of the noise source. The role of the noise memory is to induce an
enhancement of fluctuations of the number of infected people
$I(t)$ and to shift the critical value. When the noise parameter
$f$ is near the critical value with zero memory, a
stability-instability transition occurs. For epidemic dynamics we
find that, for pulse amplitude parameter less than the critical
value ($f < f_c$), we obtain quick recovering. Moreover we find
that the role of the noise memory, that is the incubation period,
is to reduce the amplitude of the fluctuations of the infected
people. This causes a more regular time behavior of the illness.

Finally concerning population dynamics of some species, this noise
can represent situations in which the species density growth is a
quick process, and the decreasing process is continuous in time.
The delay $\zeta_0$ should represent the time taken by the
individuals of the species to become adult. So we could estimate
the dependence of population dynamics on this parameter, with all
other ones fixed.

\section{Acknowledgments}

We are very grateful to professor I. Sokolov and professor Yu.
Romanovskij for useful and interesting discussions. This work has
been supported by INTAS Grant 01-450, INFM and MIUR.

\end{document}